# *SparseAssembler2*: Sparse *k*-mer Graph for Memory Efficient Genome Assembly

Chengxi Ye[1,2,7], Charles H. Cannon[3,4], Zhanshan (Sam) Ma[2,6]*, Douglas W. Yu[5,6], Mihai Pop[7]*

[1]Ecology & Evolution of Plant-Animal Interaction Group, Xishuangbanna Tropical Botanic Garden, Chinese Academy of Sciences, Menglun, Yunnan 666303 China.

[2]Ecology, Conservation, and Environment Center; State Key Laboratory of Genetic Resources and Evolution, Kunming Institute of Zoology, Chinese Academy of Sciences, Kunming, Yunnan 650223 China.

[3]Ecological Evolution Group, Xishuangbanna Tropical Botanic Garden, Chinese Academy of Sciences, Menglun, Yunnan 666303 China.

[4]Department of Biological Sciences, Texas Tech University, Lubbock, TX 79410 USA.

[5]School of Biological Sciences, University of East Anglia, Norwich, Norfolk NR47TJ UK.

[6]Computational Biology and Medical Ecology Lab; State Key Laboratory of Genetic Resources and Evolution, Kunming Institute of Zoology, Chinese Academy of Sciences, Kunming, Yunnan 650223 China.

[7]Department of Computer Science and Center for Bioinformatics and Computational Biology, Institute for Advanced Computer Studies, University of Maryland, College Park, MD, USA.

**ABSTRACT**
**Motivation:** To tackle the problem of huge memory usage associated with *de Bruijn* graph-based algorithms, upon which some of the most widely used *de novo* genome assemblers have been built, we released *SparseAssembler1*. *SparseAssembler1* can save as much as 90% memory consumption in comparison with the state-of-art assemblers, but it requires rounds of denoising to accurately assemble genomes. Algorithmetically, we developed an extension of *de Bruijn* graph structure — 'sparse *de Bruijn* graphs' — skipping a certain number of intermediate *k*-mers. In this paper, we introduce a new general model for genome assembly that uses only sparse *k*-mers. The new model replaces the idea of the *de Bruijn* graph from the beginning, and achieves similar memory efficiency and much better robustness compared with our previous *SparseAssembler1*.

**Results:** Based on the sparse *k*-mers graph model, we develop *SparseAssembler2*. We demonstrate that the decomposition of reads of all overlapping *k*-mers, which is used in existing *de Bruijn* graph genome assemblers, is overly cautious. We introduce a sparse *k*-mer graph structure for saving sparse *k*-mers, which greatly reduces memory space requirements necessary for de novo genome assembly. In contrast with the *de Bruijn* graph approach, we devise a simple but powerful strategy, *i.e.*, finding links between the *k*-mers in the genome and traversing following the links, which can be done by saving only a few *k*-mers. To implement the strategy, we need to only select some *k*-mers that may not even be overlapping ones, and build the links between these *k*-mers indicated by the reads. We can traverse through this sparse *k*-mer graph to build the contigs, and ultimately complete the genome assembly. Since the new sparse k-mers graph shares almost all advantages of *de Bruijn* graph, we are able to adapt a Dijkstra-like breadth-first search algorithm, for the new sparse *k*-mer graph in order to circumvent sequencing errors and resolve polymorphisms.

**Availability:** Programs in both Windows and Linux are available at: https://sites.google.com/site/sparseassembler/.

**Contact:** ma@vandals.uidaho.edu
or mpop@umiacs.umd.edu

---

*To whom correspondence should be addressed.

# 1 INTRODUCTION

Genome assembly is one of the few most foundational operations in bioinformatics and computational biology. The computational algorithms used for it evolve with the advances in sequencing technology. The earlier algorithms used for genome assembly with data produced by the second-generation sequencing technology such as *Roche GS FLX '454'* and *Illumina* belong to the so-termed Overlap-Layout-Consensus (OLC) approach, which analyzes the overlap graph of the reads and searches for a consensus genome. This OLC approach leads to NP-hard Hamilton path problem, and not only relies on possibly expensive heuristic algorithms but also consumes huge amount of memory. Examples of genome assembly software packages include: *ARACHNE* (Batzoglou, et al., 2002), *Phusion* (Mullikin and Ning, 2003), *Atlas* (Havlak, et al., 2004), *Celera* (Hunt, et al., 2004), or *phrap* (http://www.phrap.org). The OLC approach, when applied to the second generation short read sequencing (SRS) data, suffers from performance setbacks because too many overlaps have to be calculated.

In recent years, an alternative algorithm, which is based on the *de Bruijn* graph, to OLC approach resolves some of the computational challenges presented by SRS data. Several software packages using the *de Bruijn* graph (e.g., Birney and Zerbino, 2008; Birol, et al., 2009; Chaisson, et al., 2004; Himmelbauer, et al., 2007; Sundquist, et al., 2007; Wang, et al., 2010; Warren, et al., 2007) have been released since the first package, the *EULER* assembler, was introduced by (Pevzner, et al., 2001). The *EULER* assembler, in effect, converts the assembly problem into one of finding Eulerian paths. The *de Bruijn* graph is constructed using the unique words of $k$ nucleotides or $k$-mers (Fig. 1a). Reads are represented as paths through the graph, traversing from $k$-mer to the next in a specific order (see review by (Birney and Zerbino, 2008)).

In comparison with the earlier OLC approach based genome assemblers, the *de Bruijn* graph based assemblers avoid the NP-hard Hamilton path problem, but still suffer from the huge computational memory demands, which poses a major practical limitation even for moderate-size genomes. Although great efforts have been made by using graph simplification techniques such as combining the nodes corresponding to the forward and reverse complements and collapsing un-branched paths, the improvements from those ad-hoc tactics did not result in sufficient reductions to allow assembly of moderate genomes on typical desktop computers. Conway and Bromage (2011) recently tackled the memory usage problem by realizing that a naïve node-and-pointer *de Bruijn* graph representation takes a huge amount of memory space. They noticed that existing $k$-mers can be inferred by intersections of subsequent $(k + 1)$-mers (i.e. the edges in (Conway and Bromage, 2011)), which prompted them to build a bitmap recoding the presence/absence of the $4^{k+1}$ possible (k+1)-mers. Since the *assembly graph* is almost always a small subset of the full *de Bruijn* graph represented with the bitmap structure, it is actually sparse. The sparse bitmap representation, which was implemented with entropy-based succinct data structures (Okanohara and Sadakane, 2007) for bitmap compression and query, reduced memory requirements by a factor of ~10, compared to the naïve node-and-pointer structures predominantly used by most existing *de Bruijn* graph based packages.

Nevertheless, *SparseAssembler1* (Ye, et al., 2011) demonstrated that it is totally feasible to achieve similar magnitude memory-efficiency with the succinct data structures, adopted by Conway and Bromage (2011), in genome assembly with an easy-to-implement node-and-pointer approach. However, *SparseAssembler1* required rounds of denoising in order to preserve the accuracy of the assembly. In this paper, we will show with *SparseAssembler2* that the assembly could be safely reached without denoising at all. We demonstrate it is unnecessary to build the ***de Bruijn* graph** from the beginning and show a **sparse $k$-mer graph** is all that is necessary of a short read assembly. The sparse $k$-mer graph needs to save only sparsely selected $k$-mers and the links between them rather than saving dense overlapping $k$-mers as used in with the *de Bruijn* graph. Since the $k$-mers are much fewer and the links are cheap, we can greatly reduce the computational memory demands even with a simple node-and-pointer hash table approach.

The *SparseAssembler2* presented in this paper has a distinct strategy for achieving memory usage efficiency compared with existing *de Bruijn* graph based assemblers designed for the second generation sequencing technology, including those of (Conway and Bromage, 2011) and *SparseAssembler1 (Ye, et al., 2011).* Rather than building an extension of the de Bruijn graph and seeking efficient data structures to store the *de Bruijn* graph, we introduce a new and general **sparse $k$-mer graph.** The sparse $k$-mer graph captures the essence of a short read assembly, and denoising is no longer a necessary step. Memory requirement with the new strategy is on par with our previous *SparseAssembler1*. We also show that the sparse $k$-mer graph shares almost all advantages of the *de Bruijn* graph. For example, we can adapt a Dijkstra-like breadth-first search algorithm, to circumvent sequencing errors and resolve polymorphisms. These improvements make it possible to assemble moderate genomes robustly with small memory requirement available on PC platform. We have tested the *SparseAssembler2* with both simulated and real data and have demonstrated that ~90% memory space saving can be achieved with a practical value of $g = 25$, where $g$ is the number of intermediate $k$-mers skipped in the new assembler.

## 2 THE IDEA OF NEW SPARESE *K*-MERS GRAPH

We model short read assembly from SRS data with no errors in a more general and novel idea: finding links between the $k$-mers in the genome and then traverse. The *de Bruijn* graph structure (Pevzner, et al., 2001) is a special case of this idea, since it saves all the subsequent $k$-mers in the reads and the links between these $k$-mers in constructing the graph. The graph is therefore a full (with respect to the reads) overlapping $k$-mer graph, and direct implementations of the graph consume huge amounts of memory even with moderate-size genomes. In contrast with the *de Bruijn* graph approach, we realize that the novel idea can be formulated as a general algorithm by saving only a few $k$-mers, and, in effect, we are building a sparse $k$-mers graph. To realize the idea, we need to only select some $k$-mers which might not even be overlapping ones, and build the links between these $k$-mers indicated by the reads. Then, we can traverse in this sparse $k$-mer graph to build the contigs.

## 3 METHODS AND IMPLEMENTATION

### 3.1 The *de Bruijn* graph

In genome assembly, a *de Bruijn* graph structure is built from nodes of all unique length-*k* fragments, or *k*-mers, of a genome (e.g., Birney and Zerbino, 2008; Pevzner, et al., 2001). If two *k*-mers overlap by *k*-1 length, there is a directed edge from the first *k*-mer node to the succeeding one with a corresponding edge in the reverse direction. Multiple edges arise if one *k*-mer overlaps with multiple, different *k*-mers by *k*-1 length on the side.

Edges can be implicitly represented by saving only the presence information of the neighboring nucleotides. A common first stage of *de Bruijn* graph based *de novo* assemblers is to build the graph by storing all the *k*-mers and their neighboring nucleotide(s). A *k*-mer is considered being different only in orientation with its reverse complement, and only one of the two (chosen by lexical-order) is saved. Let all *k*-mers be encoded in bits: 00, 01, 10, 11, respectively, for A, C, G, T, and let 4 bits be used to indicate the presence/absence of the 4 possible edges/nucleotides on every side (Fig. 1a,b). Thus, each *k*-mer uses $2 \times k + 4 \times 2$ bits of memory, and the minimum space requirement $S_1$ for a genome with *k*-mer diversity $N$ is approximately $S_1 = N \times (2 \times k + 4 \times 2)$, assuming no sequence redundancies, errors and branches. For real world situations, the requirement can be much greater (interested readers may refer to the online memory estimator for *Velvet*).

Typically, *k*-mer sizes of 21~51 bp are used because short *k*-mers result in branching, and therefore, in ambiguity in the assembly. As a consequence, the memory space required for saving all *k*-mers can be huge. Over 100 GB memory space usages are common examples to assemble the genome of many species (Wang, et al., 2010).

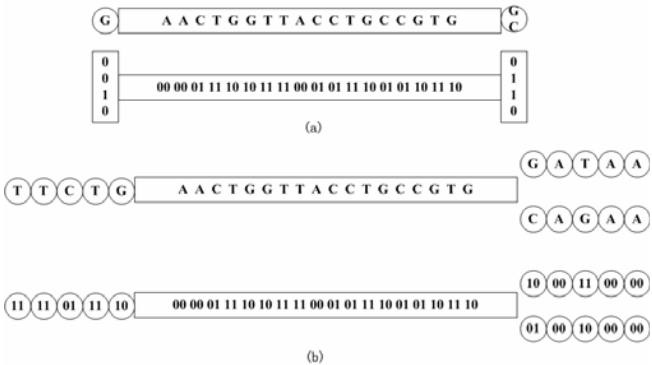

**Fig. 1.** A node with branches in the *de Bruijn* graph and the sparse *k*-mer graph. (a) A node with branches in a *de Bruijn* graph. (b) The binary implementation of (a). (c) A node with branches in a sparse *k*-mer graph. (d) The binary implementation of (c). The *k*-mers which are nodes in the graph are squared in the blocks. Neighboring nucleotides indicating the edges of the graph are circled.

### 3.2 The sparse *k*-mer graph

In the sparse *k*-mer graph structure, every node in the graph is expected to be sparsely spaced and differs by having longer links, or having more nucleotides per branch (Fig. 1c). With the sparsely spaced nodes, the memory requirement for constructing the sparse *k*-mers graph can be considerably less than that for building *de Bruijn* graphs.

Let *g* be the number of skips between *k*-mers. In an ideal case with no branch and assuming that the *k*-mers are staggered by $g = 5$ bases (we use $g = 16\text{-}25$ in our implementation), we can store $\leq 5$ neighboring bases on each side of the *k*-mer, which requires $2\times5$ bits for each side of the *k*-mer. However, we need to store much fewer *k*-mers than the existing approaches. For example, if $g = 5$, only every fifth *k*-mer needs to be stored in a hash table, and the total memory requirement for the *k*-mers is reduced to near 1/5 of that required by *de Bruijn* graph structure. Of course, for larger *g's*, the memory requirement drops even more accordingly. The real reduction is, however, somewhat less than this, for reasons to be explained in the next subsection.

It is interesting to note the low bound of memory space usage for the sparse *k*-mer graph. Assume that the saved *k*-mers are all *g* bases apart. We reduce the number of stored *k*-mers to a fraction of $1/g$. So the total memory space requirement is $S_2 \approx \frac{N}{g} \times (2 \times k + 2 \times 2 \times g + \text{ptr\_sz}) = N \times (\frac{2 \times k}{g} + 2 \times 2 + \frac{\text{ptr\_sz}}{g})$, in which *ptr_sz* is the extra space required by the pointers structures for the edge links. Relative to the *de Bruijn* graph, we reduce *k*-mer storage to $1/g$, and the portion for storing edges to a half, but add a new space requirement for storing edge links. Let reads be of length *r*; the sparse *k*-mers scheme becomes more efficient when *r-k* is large. Therefore we can use large *g*'s and still get informative reads, which becomes more common with the improvements of future sequencing technology. Following this trend of technology advances, we can increase *k* to larger values than those used by previous assembly methods while still keeping memory usage low.

### 3.3 Building the sparse *k*-mer graph

We build a graph with the sparse *k*-mers in two rounds. In the first round, we select the *k*-mers that will be used as the nodes. For a preset *g*, we scan each read and see if any of the subsequent *g* *k*-mers are already used as a new node. If so, we move to that new node and continue the scan. Otherwise, we select the current *g* *k*-mers as a node. After the first scan, the nodes are selected, and they are expected to be nearly *g*-gapped if there are no sequencing errors. In real data, we filter off the lowly covered nodes before we move to the second round. Lowly covered nodes are regarded as nodes in spurious branches such as tips or bubbles or a real *k*-mer node connected to a spurious branch. In the second round, the links between the selected nodes are built. The accurate coverage of the *k*-mer nodes are recalculated in this round. After the two rounds of processing, the *k*-mers picked as nodes are expected (*i.e.*, on average), but not strictly, *g*-gapped, which results in redundancy in space. The sparse *k*-mer graph is well defined and built via the above-described two-rounds processing with all the reads, but the real *de Bruijn* graph is never stored!

### 3.4 Circumventing sequencing errors and graph simplification

Sequencing errors and polymorphisms can result in tips or bubbles (Birney and Zerbino, 2008) in *de Bruijn* graph, which is also the case in the sparse *k*-mer graph. To remove these unwanted structures, we first remove the weak links after round 2. After that, like in *Velvet* (Birney and Zerbino, 2008), we developed a Dijkstra-like breadth-first search algorithm to further detect unwanted structures. The search backtracks to the last branching node upon reaching a visited node or a tip end. We choose the more heavily covered branch and remove the tips. After this, spurious paths and redundant structures like tiny loops and bubbles are removed (Fig. 2).

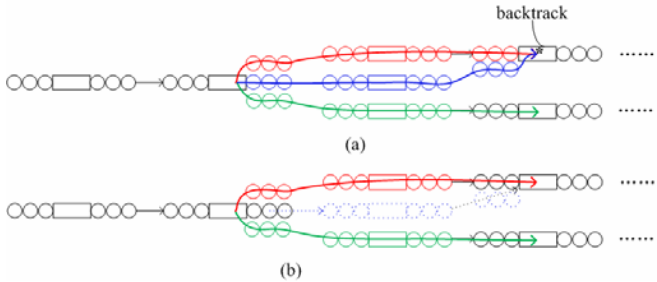

**Fig. 2.** Breath-first search bubble removal. Removing unwanted structures in the sparse *de Bruijn* graph. (a) Before removal. (b) After removal.

### 3.5 Genome assembly

The full assembly process consists of (*i*) building the sparse *k*-mer graph with the sparse *k*-mer nodes identified during the 2 rounds of processing described above and (*ii*) graph traversal. The procedure for reconstructing the genome is similar with that used by the *de Bruijn* graph algorithms. A new traversal begins at a node not visited in previous traversals, and breaks when branches are detected; the separate traversals form the contigs.

## 4 RESULTS

In the noise-free simulation, our sparse k-mers based *SparseAssembler2* consistently outperformed several state-of-the-art assemblers with the fruit fly (Table 1), rice (Table 2), and *E. coli* (Tables 3 & 4) genomes, including *SOAPdenovo*, *Velvet* and *ABySS*; the new assembler uses substantially less computational memory in comparable job times. For the fruit fly, the longest contig and N50 were substantially longer than any other assembler. While the results for the rice genome were roughly equivalent, our N50 was better. In all the comparisons, we simulated length 100 reads and used $k = 31$, $g = 25$. For the fruit fly and rice genomes, the comparisons did not include any error, because the existing standard assemblers required too much memory to run on our desktop computing environment, but the comparison on the *E. coli* dataset was performed with and without errors (Table 3 & 4). For the *E. coli* genome, we simulated substitution errors using *MetaSim* (Richter, et al., 2008), with growing error rates from 0.5% (5' end) to 2% (3' end). More than 2 million errors were added by *MetaSim* to the 3 million simulated reads, which heavily exacerbated the memory burden. Our *SparseAssembelr2* can still get reasonable results with the highest contig N50 (Table 4).

To test performance on real data, we compared our approach on 100 bp read whole genome shotgun sequence data generated on the *Illumina* platform for the 416 Mbp bee (*Lasioglossum albipes*) genome with 60X coverage (D. Yu, unpublished dataset). $k = 31$ was used for both assemblers, and we set $g = 25$ for *SparseAssembler2* (Table 5). Our approach was substantially more efficient than *SoapDenovo*, in terms of memory usage (3.5 GB vs. ~30 GB), and obtained a slightly better assembly in terms of both N50 (1790 vs. 1689) and total assembled length (262 Mbp vs. 258

Mbp). Our time consumption on this large real dataset was also smaller.

We have not explored efficient implementations of the data structures previously used in SparseAssembler1 (Ye et al. 2011). For example, lots of pointers used in the hash table are relative expensive; some of which are unnecessary and will be omitted in future implementations. We also noticed the runtime memory is at least 20% higher than the real memory needed in this implementation. We will resolve these minor issues in the next version, which should further improve the performance of *SparseAssembler2*.

**Table 1. Assembly performance comparison on the fruit fly genome**

|  | *ABySS* | *Velvet* | *SOAPdenovo* | *SparseAssembler2* |
|---|---|---|---|---|
| Time (hr) | 1.5 | 1.5 | 0.5 | 0.8 |
| Memory peak (GB) | 6.2 | 8 | 4.5 | 0.7 |
| Longest contig (bp) | 162,263 | 190,106 | 200,772 | 273,977 |
| >10 kbp (# contigs) | 3,368 | 3,266 | 3,098 | 2,794 |
| Sum (kbp) | 82,175 | 87,758 | 91,843 | 97,482 |
| >100 bp (# contigs) | 23,981 | 22,035 | 20,394 | 19,360 |
| Sum (kbp) | 113,564 | 113,642 | 113,683 | 113,926 |
| Mean size (bp) | 4,736 | 5,158 | 5,574 | 5,885 |
| N50 (bp) | 19,893 | 24,546 | 29,314 | 38,835 |
| Coverage (%) | 94.41 | 94.47 | 94.51 | 94.71 |

The performance on the fruit fly genome dataset, genome size: 120,291 kbp. Programs are run on default settings. The mean size, N50, Coverage in the last three rows are calculated based on the contigs longer than 100 bp. The fruit fly genome is obtained from GenBank: X, NC_004354.3; IIL, NT_033779.4; IIR, NT_033778.3; IIIL, NT_037436.3; IIIR, NT_033777.2; IV, NC_004353.3

**Table 2. Assembly performance comparison on the rice genome**

|  | *ABySS* | *Velvet* | *SOAPdenovo* | *SparseAssembler2* |
|---|---|---|---|---|
| Time (hr) | 5 | 5 | 1.5 | 2.7 |
| Memory peak (GB) | 15.5 | 30.0 | 6.3 | 2.0 |
| Longest contig (bp) | 23,220 | 26,881 | 26,869 | 27,077 |
| >10 kbp (# contigs) | 461 | 527 | 656 | 1,019 |
| Sum (kbp) | 5,683 | 6,507 | 8,161 | 12,823 |
| >100 bp (# contigs) | 459,438 | 402,431 | 454,110 | 460,048 |
| Sum (kbp) | 254,793 | 227,835 | 261,911 | 277,868 |
| Mean size (bp) | 555 | 566 | 577 | 568 |
| N50 (bp) | 1,434 | 1,516 | 1,593 | 1,784 |
| Coverage (%) | 68.73 | 61.31 | 70.48 | 74.95 |

The performance on the rice genome dataset, genome size: 370,733 kbp. Programs are run on default settings. The mean size, N50, Coverage in the last three rows are calculated based on the contigs longer than 100 bp. The rice genome is obtained from the IRGSP website (http://rgp.dna.affrc.go.jp/J/IRGSP/Build3/build3.html).

**Table 3. Assembly performance on the *E. coli* genome (noise free)**

|  | ABySS | Velvet | SOAPdenovo | SparseAssembler2 |
|---|---|---|---|---|
| Time (min) | 3 | 2 | 0.5 | 0.8 |
| Memory peak (MB) | 277 | 421 | 321 | 30 |
| Longest contig (bp) | 127,976 | 127,976 | 128,055 | 138,264 |
| >10 kbp (# contigs) | 146 | 149 | 145 | 142 |
| Sum (bp) | 3,528,911 | 3,592,598 | 3,615,431 | 3,839,644 |
| >100 bp (# contigs) | 543 | 536 | 527 | 520 |
| Sum (bp) | 4,530,909 | 4,547,178 | 4,547,902 | 4,552,397 |
| Mean size (bp) | 8,344 | 8,484 | 8,630 | 8,755 |
| N50 (bp) | 22,173 | 23,326 | 23,970 | 28,478 |
| Coverage (%) | 97.66 | 98.01 | 98.01 | 98.11 |
| E≥1* | 4 | 2 | 8 | 0 |
| E≥3 | 1 | 2 | 8 | 0 |
| E≥5 | 1 | 1 | 5 | 0 |

\* Records the number of contigs that contain more errors than the thresholds (1, 3, 5). Errors are found by base-wise comparison of the assembled contigs (with length > 100) with the ground truth genome.

**Table 4. Assembly performance on the *E. coli* genome with simulated errors by *MetaSim*.**

|  | ABySS | Velvet | SOAPdenovo | SparseAssembler2 |
|---|---|---|---|---|
| Time (min) | 37 | 5.5 | 1.5 | 2 |
| Memory peak (GB) | 4.1 | 2.4 | 1.9 | 0.2 |
| Longest contig (bp) | 127976 | 120,922 | 127,978 | 128,055 |
| >10 kbp (# contigs) | 147 | 146 | 150 | 145 |
| Sum (bp) | 3,543,016 | 3,486,383 | 3,618,828 | 3,632,561 |
| >100 bp (# contigs) | 544 | 550 | 538 | 529 |
| Sum (bp) | 4,545,014 | 4,544,303 | 4,545,879 | 4,541,646 |
| Mean size (bp) |  | 8,262 | 8,450 | 8,585 |
| N50 (bp) | 22173 | 22173 | 23,336 | 24,740 |
| Coverage (%) | 97.96 | 97.94 | 97.98 | 97.89 |
| E≥1 | 5 | 62 | 11 | 4 |
| E≥3 | 2 | 24 | 7 | 2 |
| E≥5 | 1 | 15 | 7 | 1 |

Three million reads are simulated with MetaSim, with error rates increasing from 0.5% (5' end) to 2% (3' end). Programs are run on tuned parameters. In *SOAPdenovo*, we set -d 2 -D2.

**Table 5. Assembly performance comparison on the *Lasioglossum albipes* genome (416 Mbp).**

|  | SOAPdenovo | SparseAssembler2 |
|---|---|---|
| Time (hr) | 6.7 | 4.0 |
| Memory peak (GB) | 27 | 3.5 |
| Longest contig (bp) | 27,079 | 35,275 |
| >10 kbp (# contigs) | 462 | 564 |
| Sum (kbp) | 5,778 | 7,198 |
| >100 bp (# contigs) | 335,218 | 341,547 |
| Sum (bp) | 258,324 | 262,374 |
| Mean size (bp) | 771 | 768 |
| N50 (bp) | 1,689 | 1790 |

## 5 CONCLUSION AND DISCUSSION

The above comparative analysis of *SparseAssembler2* with some state-of-the-art *de Bruijn* graph based assemblers demonstrated that our sparse *k*-mer graph, as an alternative to the sparse *de Bruijn* graph, is not only feasible for genome assembly with SRS data generated from the second generation sequencers, but also significantly and consistently outperforms some of the best-performed existing assemblers in terms of the memory usage efficiency. The *SparseAssembler2* also achieves similar or slightly better performance in terms of other metrics such as N50 and maximum contig lengths, compared with existing assemblers. This indicates that it is totally feasible to perform *de novo* genome assembly on PC platform with our sparse *k*-mers graph based approach such as implemented with *SparseAssembler2*.

The memory savings achieved by *SparseAssembler2* is similar to that achieved with Conway & Bromage's succinct data structure (Conway and Bromage, 2011) and *SparseAssembler1* (Ye, et al., 2011). But the *SparseAssembler2* does not require any denoising, and is simpler in idea and implementation. Furthermore, the saving of our assemblers is scalable with the length of *g*, which is consistent with the improvement trend of current sequencing technology, *i.e.*, increasing *g* length. Finally, the sparse k-mers graph shares almost all advantages of the sparse *de Bruijn* graph model.

Recent comparative studies conducted by Deng, et al., 2011; Zhang, et al., 2011 on several existing *de novo* assembly packages, including *SSAKE*, *VCAKE*, *Euler-sr*, *Edena*, *Velvet*, *ABySS* and *SOAPdenovo*, failed to discover significant differences in the magnitude of the memory usages which were all large among packages. Nor were major performance differences found between simulated and real data. These comparative studies therefore suggest that comparisons with only a few other assembly packages, rather than all existing ones, should be sufficient in order to gauge the relative performance of our approach. This justifies our selective comparisons with only 3 major state-of-the-art assemblers, *ABySS*, *Velvet* & *SOAPdenovo*, on the *E. coli*, fruit fly, rice genomes and a real bee genome (Table 1-5), and our approach consistently consumed much less memory space. Therefore, the results reported here prove our idea that a sparse *k*-mer graph retains sufficient information for accurate and fast *de novo* genome assembly in a cheap, desktop PC computing environment, which is usually only equipped with several gigabytes memory and the cost can be ignored compared with currently used super computers. Future improvements to *SparseAssembler2* will focus on exploitation of paired-end reads.

## ACKNOWLEDGEMENTS

We thank Prof. Jin Chen of the Xishuangbanna Tropical Botanical garden, Dr. Jue Ruan of the Beijing Institute of Genomics and Hao Tan of BGI for long-term support and fruitful discussions. The *Lasioglossum albipes* genome data were generously provided by DY's collaborators on this project: Sarah Kocher, Naomi Pierce, and Hopi Hoekstra.

*Funding*: We also acknowledge support from Yunnan Province (20080A001), China and the Chinese Academy of Sciences (0902281081, KSCX2-YW-Z-1027, Y002731079).